\documentstyle{lmamult_pre}
\begin{document}
 
\def\etal{{\it et~al.\ }}
\def\exdent#1\par{\noindent\hang\frenchspacing#1\par}
\def\plotfiddle#1#2#3#4#5#6#7{\centering \leavevmode
\vbox to#2{\rule{0pt}{#2}}
\includegraphics{#1}}

\title{The Tully-Fisher Relation at Intermediate Redshifts}
\author{Matthew A. Bershady\inst{1}\inst{2}}
 
\institute{Penn State University, State College PA 16802, USA
\and
Hubble Fellow}
 
\maketitle

\setlength{\unitlength}{1in}
\begin{picture}(5,0)(0,-3)
\put(-0.6,0.0){\makebox(5,0){Proceedings of ``Galaxies in
the Young Universe,'' 1996, Springer series}}
\put(-0.8,-0.15){\makebox(5,0){``Lecture Notes in Physics'', eds. H. Hippelein, K. Meisenheimer,}}
\put(-2.15,-0.3){\makebox(5,0){H.-J. R\"{o}ser, p.139}}
\end{picture}

\section{Introduction}

Motivated by recent pioneering measurements of galaxy kinematics at
intermediate redshifts$^{1,2}$, we have begun a pilot survey to
measure rotation curves for blue, star-forming galaxies between
0.05$<$z$<$0.35. The scientific impetus for such measurements is to
construct internal velocity - luminosity relations at redshifts
substantial enough to make cosmological and evolutionary
tests.$^{3,4}$ Departures from a fiducial relation as a function of
redshift are sensitive to both space-time curvature and the evolution
of the mass-to-light ratio in galaxies. Because of this ambiguity --
curvature vs. evolution -- kinematics should be viewed as a new
cosmological tool at intermediate redshifts to be combined with
additional measurements. In particular, the apparent internal
velocities, colors, surface-brightness, and image shape are all
unaffected by curvature, so that this ensemble of observables can be
used unambiguously to explore galaxy evolution.

Here we discuss a well-defined method of target selection which
assures efficient measurement of spatially and spectrally resolved
kinematics, namely rotation curves. Spatial resolution is critical
since it is difficult to use integrated line-widths to distinguish
between, for example, a low-mass star-forming galaxy and a high-mass
galaxy with a central star-burst. For 18 of 19 appropriately selected
galaxies to $B$$<$20.5 from $[5]$, we have successfully measured
rotation curves using the KAST spectrograph on the Lick 3m telescope
with 1-2 hour integrations for each target. As a sanity check, we
present preliminary results for half of our sample: We measure H$_0$
to be $\sim$ 75 km s$^{-1}$ Mpc$^{-1}$ at a median redshift of 0.15.

\begin{figure}
\plotfiddle{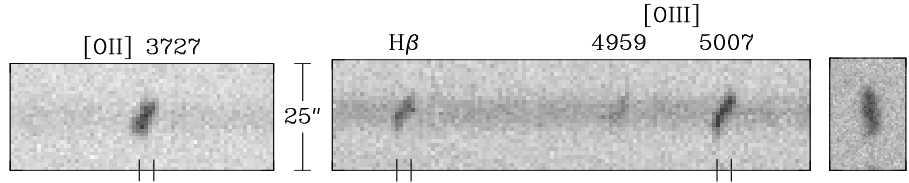}{2.5cm}{270}{50}{50}{-193}{220}
\caption{An example of an observed rotation-curve spectrum for one of
our higher-redshift targets at z=0.3 (sa68.5154 from $[5]$ and $[6]$)
using the KAST spectrograph and Lick 3m telescope. We derive a maximum
rotation velocity of 405$\pm$20 km/s, peak-to-peak from [OII],
H$\beta$, and [OIII]5007. [OIII]4959 suffers from being partly on a
sky-line. [OII] appears fuzzy because we have not resolved this
doublet. A gound-based $B$ band image of the galaxy (to scale), taken
in $\sim$ 1'' FWHM seeing at the KPNO 4m, is at the far right.}
\end{figure}

\section{Target Selection}

There are three desiderata for selecting galaxies for rotation-curve
surveys at substantial redshift: (i) A similar range of galaxies
should be chosen at disparate redshifts on the basis of some objective
criteria. (ii) Targets should be optimized for the available
instrumental resolution. (iii) Targets should be efficient to
observe. There are several well-known correlations for galaxies that
when combined, point to a well-defined galaxy type. The correlations
are between size and luminosity, color and emission-line strength,
color and luminosity, and luminosity and internal velocity. These
conspire to make the ideal rotation-curve targets blue, luminous
galaxies. How luminous depends on the spatial and spectral resolution
of your spectrograph and telescope, and the redshift limit of your
survey. How blue depends on your desired efficiency. For our sample,
we were able to measure rotation curves one at a time on a 3m
telescope at a pace that outstriped the multiplexing advantage of
observing red galaxies in clusters at comparable redshifts using a
bigger telescope.$^1$ At the same time, we could spectrally and
spatially resolve rotation curves to z=0.3 (see fig. 1) with 1-2''
seeing and R$\sim$1500-2500 in the red.

The ideal galaxy for intermediate and high redshift rotation curve
studies has spectral type ``bm'' in the nomenclature of $[6]$. This is
comparable to luminous ``Sc'' galaxies. However for the purposes of
selecting these galaxies for surveys on 4m-class telescopes, it is
much more fruitful to think in terms of spectral types. Depending on
how restrictive one makes the above selection, the surface-density of
luminous, blue galaxies to $B$=20.5 is low. Without substantially
increasing telescope aperture as well as spatial and spectral
resolution, fainter limits will not produce substantially greater
surface-densities of viable targets: Fainter galaxies will either be
at higher redshift (and apparently smaller), or lower in luminosity
(and internal velocity and apparent size) at comparable
redshifts. This makes selection via Hubble type rather inefficient
since this requires Hubble Space Telescope (HST) images which cover
little area. In contrast, ground-based, multi-band imaging can provide
photometric redshifts$^7$ and hence spectral classification and
luminosities over large areas, ideal for selecting targets for
spectroscopic follow-up and pointed HST imaging (when needed). With
adaptive optics and/or a 10m-class telescope, target selection
strategy can be altered. Here one is optimally exploring a different
redshift and/or luminosity regime than should be pursued with a
conventional 4m-class telescope.

\begin{figure}
\plotfiddle{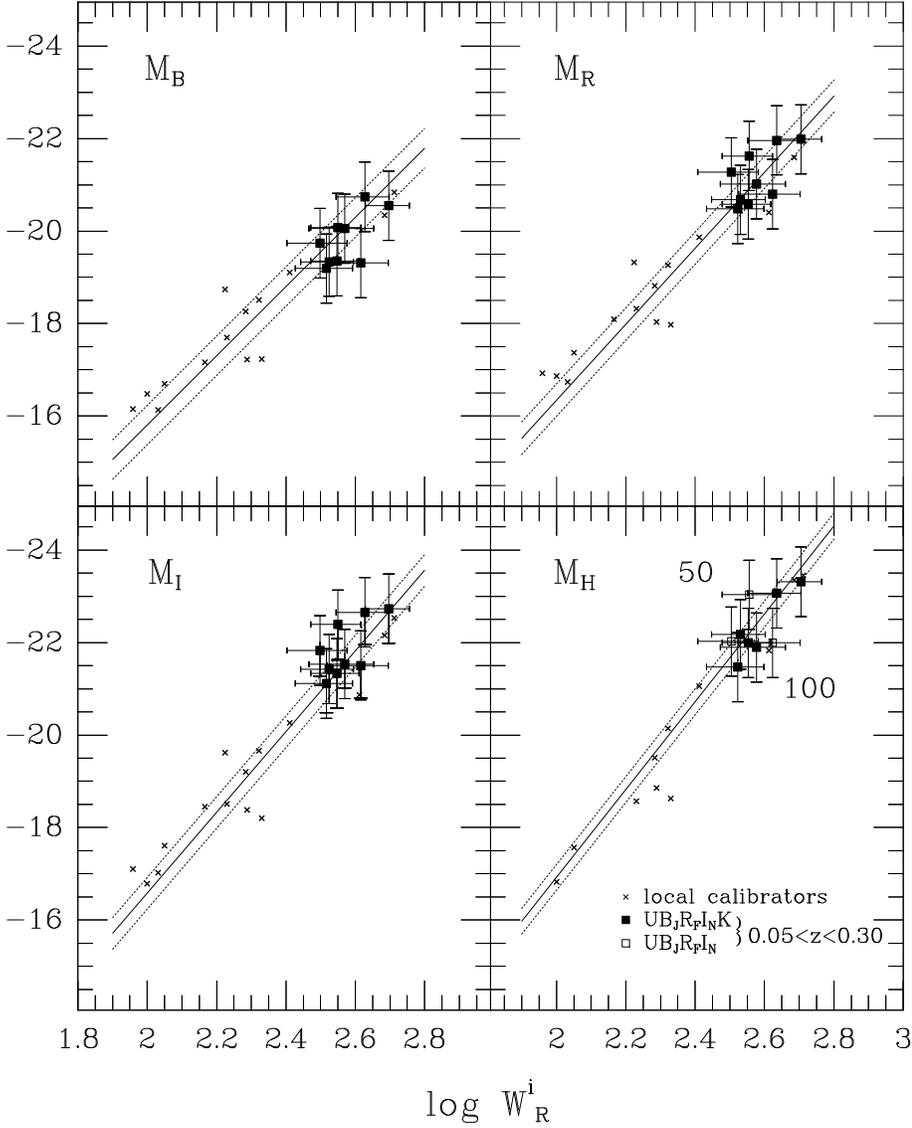}{14cm}{0}{72}{72}{-215}{-65}
\caption{The Tully-Fisher relation as derived for half our current
sample (fully processed). Local calibrating galaxies from $[9]$, with
their regression slopes and dispersions are plotted for
comparison. Transformation from rotation speed to 21-cm line-widths
for our data use formulae in $[8]$. ``Error-bars'' in the y-dimension
indicate not photometric uncertainties but different assumed values of
H$_0$, as indicated (q$_0$=0.1). Galaxies in our sample represented by
filled boxes have 5-band photometry including the $K$ band; open boxes
represent (two) galaxies without $K$ band photometry. Absolute
magnitudes in the $B$, $R$, $I$, and $H$ bands are determined
empirically by interpolating between observed bands via stellar
synthesis models fit directly to the data. $^6$}
\end{figure}

\section{A Sanity Check}

To see whether it is possible to make sensible kinematic measurements
at intermediate redshifts, we have transformed emission-line maximum
rotation velocities, measured now for half our sample, into 21-cm
line-widths using transformations from $[8]$. In fig. 2 we compare our
sample to the local calibrators of the Tully-Fisher relation from
$[9]$. The errors-bars in the x-axis (velocity) are indeed estimates
of measurement errors (rotation velocity, inclination, and
transformation uncertainties). However, in the y-axis, the photometric
uncertainties are negligible ($<$10\%, since we have deep, multi-band
photometry from which we empirically derive $\kappa$-corrections).
Instead, these error bars represent a range of {\it assumed} values of
H$_0$ between 50 and 100 km s$^{-1}$ Mpc$^{-1}$. Relative to the local
calibrators, we find average offsets indicating values between 60 and
90 km s$^{-1}$ Mpc$^{-1}$, depending on band. The median redshift of
this sample is 0.15. This result is preliminary since we haven't, for
example, corrected for internal extinction (although the correction is
negligible in the H band), or corrected for more subtle differences
between our photometric system and that used in $[9]$. Furthermore,
inclination corrections have been made on the basis of ground-based
images and should be checked with higher-resolution data.

Future prospects should be even more promising. The measurement we are
actually interested in making is {\it differential}, and not
absolute. That is, we want to study the deviations from, or changes in
scatter about {\it some} fiducial correlation of internal velocity and
luminosity as a function of redshift. Instead of using integrated
line-widths or transformations to another system imposed by
observations of local galaxies, we can instead devise a new standard
measurement of internal velocity (and luminosity) that is optimal for
intermediate to high redshift studies. One possibility along these
lines has already been suggested.$^{10}$

\medskip

The work presented here involves a collaboration with D. Koo and
C. Mihos (Lick Observatory). MAB was funded by NASA through grant
HF-1028.02-92A from STScI (operated by AURA, Inc. under contract
NAS5-26555).

\end{document}